\documentclass[a4paper]{article}

\usepackage{itgspeech2023}    
\usepackage{times}            
\usepackage[english]{babel}   
\usepackage[ansinew]{inputenc}
\usepackage[T1]{fontenc}      
\usepackage[sort&compress,numbers]{natbib}	
\usepackage{amsmath,amssymb}
\usepackage{graphicx}
\usepackage[colorlinks=false,pdfborder={0 0 0}]{hyperref}
\usepackage{units}
\usepackage{hyperref}
\usepackage{subcaption}
\usepackage{graphicx}
\usepackage{multirow}
\usepackage{booktabs}

\title{In-the-wild Speech Emotion Conversion Using Disentangled Self-Supervised Representations and Neural Vocoder-based Resynthesis}

\author{Navin Raj Prabhu$^{\star \dagger}$, Nale Lehmann-Willenbrock$^{\dagger}$, Timo Gerkmann$^{\star}$ }

\address{$^{\star}$Signal Processing, $^{\dagger}$Industrial and Organizational Psychology, Universit\"at Hamburg, Germany\\
  Email: \texttt{\{navin.raj.prabhu, nale.lehmann-willenbrock, timo.gerkmann\}@uni-hamburg.de}}

\begin{document}

\maketitle

\begin{abstract}
Speech emotion conversion aims to convert the expressed emotion of a spoken utterance to a target emotion while preserving the lexical information and the speaker's identity. In this work, we specifically focus on in-the-wild emotion conversion where parallel data does not exist, and the problem of disentangling lexical, speaker, and emotion information arises. In this paper, we introduce a methodology that uses self-supervised networks to disentangle the lexical, speaker, and emotional content of the utterance, and subsequently uses a HiFiGAN vocoder to resynthesise the disentangled representations to a speech signal of the targeted emotion. For better representation and to achieve emotion intensity control, we specifically focus on the aro\-usal dimension of continuous representations, as opposed to performing emotion conversion on categorical representations. We test our methodology on the large in-the-wild MSP-Podcast dataset. Results reveal that the proposed approach is aptly conditioned on the emotional content of input speech and is capable of synthesising natural-sounding speech for a target emotion. Results further reveal that the methodology better synthesises speech for mid-scale arousal (2 to 6) than for extreme arousal (1 and 7).
\end{abstract}

\section{Introduction}
One way in which emotions are expressed by individuals in social interactions is via speech signals \cite{schuller2018speech}. In the context of human-machine interaction systems, the generation of spoken dialogue is a fundamental facet of natural interaction between humans and machines \cite{crumpton2016survey, zhou2022mixedemo}. More importantly, to improve the naturalness of machine communication, the generation of emotionally expressive speech is required. While speech generation technologies have been making significant progress \cite{hifigan, hsu2023revise}, emotionally expressive speech generation is still a challenge \cite{triantafyllopoulos2023overview, zhou2022emotional}. Speech emotion conversion (SEC) is a technique that aims to convert the expressed emotion of a spoken utterance to a target emotion while preserving the lexical and the speaker information \cite{VCdu22c_interspeech, triantafyllopoulos2023overview}. Therefore, SEC has a crucial application in building next-generation human-machine interaction systems, aiming at equipping them with the ability to interact with social and emotional intelligence.

Emotion can be represented either \textit{categorically} or as \textit{dimensional} representations. As \textit{categorical} representations, Ekman's six basic emotions \cite{ekman1971constants} (e.g., anger, happy) are commonly used \cite{busso2008iemocap}. However, emotion is a fuzzy construct with \textit{fuzzy} class boundaries \cite{russell1980circumplex}, and such discrete representations do not aptly capture the subtle difference between human emotions \cite{russell1980circumplex}. To overcome this, the circumplex model \cite{russell1980circumplex} captures emotional expressions using two continuous and independent dimensions, i.e., \emph{arousal} (relaxed or passive vs. aroused or activated) and \emph{valence} (positive vs. negative) \cite{martinez2020msp, prabhu22_interspeech}. In SEC research, efforts have also been made to control the intensity of categorical emotion representations, e.g., using mixed emotion representations \cite{zhou2022mixedemo} or modeling emotion intensity as an auxiliary task \cite{zhou2023_emointensitycontrol}. Note that SEC using the dimensional representations directly archives intensity control, as opposed to an additional effort in the categorical representation case.

\begin{figure}[t!]
	\centerline{\includegraphics[width=.95\columnwidth]{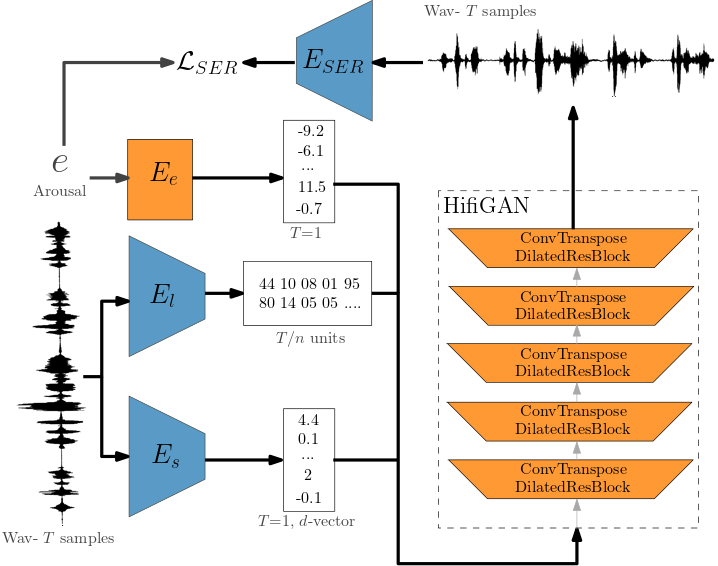}}
	\caption{Illustration of the training process. During inference, $e$ is replaced by the target emotion $\Bar{e}$. \textit{Orange} blocks denote trainable blocks, and \textit{blue} blocks are pretrained.}
	\label{fig:emovc_architecture}%
\end{figure} 

SEC datasets are primarily \textit{acted-out} \cite{busso2008iemocap, vawganzhou2021seen, zhou2022emotional, burkhardt2005database}, as opposed to \textit{in-the-wild} improvised datasets \cite{martinez2020msp}. Acted-out databases make a strong assumption on the availability of parallel utterances \cite{triantafyllopoulos2023overview}, i.e., one where each source utterance also has a ground-truth utterance of a target emotion, which is resource inefficient to collect \cite{rizos2020stargan, zhour2020CycleGAN} and techniques relying on them lack scalability \cite{triantafyllopoulos2023overview}. 

In this work, we specifically focus on non-parallel data. Models trained on non-parallel data are scalable to different emotion types \cite{triantafyllopoulos2023overview}, as they are not restricted by the emotion pairs trained on. However, they are also more challenging than modeling parallel data, as the problem of \emph{disentanglement} arises \cite{triantafyllopoulos2023overview, VCdu22c_interspeech}. For SEC on non-parallel data, a disentanglement method is required to decompose the input speech signal into several constituents: emotion, lexical, and speaker information, encoded in respective latent representations. During inference, the latent representation encoding emotion is modified to achieve emotion conversion. Existing works have primarily employed variational auto-encoders (VAE) \cite{vawgan_base, vawganzhou2021seen} and sequence-to-sequence encoder-decoders \cite{zhou2023_emointensitycontrol, zhou2022mixedemo} to achieve disentanglement. 

Self-supervised learning (SSL) techniques that leverage large unlabeled datasets have shown to be promising in several tasks, most importantly its state-of-the-art performance in speech emotion recognition (SER) \cite{deoliveira2023leveraging, wagner2023dawn} and speaker voice conversion \cite{polyak21_interspeech_resynth}. Motivated by this, in this paper, we propose a novel methodology that employs SSL models to disentangle emotion, lexical, and spe\-aker identity representations, and uses a HiFiGAN \cite{hifigan} to resynthesise the disentangled representations into emotional spe\-ech. For a better representation of emotion and better intensity control in emotion conversion, in this paper, we will represent emotion using the continuous arousal dimension. Moreover, we train and validate the proposed methodology on the in-the-wild MSP-Podcast v1.10 dataset \cite{martinez2020msp}. To the best of our knowledge, this work is the first in the literature to perform SEC on an in-the-wild setting and on arousal, while existing literature has primarily focused on categorical emotion (e.g., anger, sad, neutral, happy) \cite{vawganzhou2021seen, rizos2020stargan, zhour2020CycleGAN} and acted-out datasets (e.g., IEMOCAP, ESD) \cite{busso2008iemocap, zhou2022emotional}.


\section{Background}\label{sec:background}

\subsection{Disentangled SSL representations}\label{sec:ssl-disentangle}
SSL techniques have shown great success in several downstream tasks such as automatic speech recognition \cite{baevski2020wav2vec}, phoneme segmentation \cite{kreuk20_interspeech}, speaker verification \cite{Chen2021WavLM}, and SER \cite{deoliveira2023leveraging, wagner2023dawn}. Existing literature well documents that representations obtained from SSL models finetuned on a specific task have exclusive information on that particular task. Analysis presented in \cite{dischubertAnalysis2023} reveals that discrete representations obtained from HuBERT contain exclusively phoneme and lexical information, without any speaker or F0 information. In \cite{Chen2021WavLM}, the HuBERT framework was used to build the WavLM which was then fine-tuned exclusively to preserve speaker identity \cite{Chen2021WavLM}.
Motivated by this, in this paper, we use SSL networks to disentangle lexical and speaker information from input speech. 

\subsection{Speech resynthesis}\label{sec:speech-resynth}

While disentanglement is only the first step of SEC, the subsequent step is to {synthesise} the disentangled SSL representations into natural-sounding speech that is also conditioned on the emotion content. Traditionally, neural network based vocoders conditioned on the \text{log} \textit{Mel-spectogr\-am} enabled generating natural-sounding speech \cite{wavnet}. More recently, techniques have been proposed to directly synthesise natural-sounding speech from SSL-based speech representations \cite{polyak21_interspeech_resynth, hsu2023revise, kreuk20_interspeech, dischubertAnalysis2023}, the technique termed as \textit{resynthesis}. In \cite{polyak21_interspeech_resynth}, HiFiGAN was used to resynthesise speech from disentangled representations obtained from three SSL models: HuBERT for lexical content, VQ-VAE for F0 information, and a speaker verification model. In this paper, we follow a similar approach, that is to perform resynthesis by training a HiFiGAN on disentangled SSL representations. The crucial difference between our work and \cite{polyak21_interspeech_resynth} is that we employ this technique in SEC, thereby also requiring us to condition the HiFiGAN on emotion information.


\section{Proposed Methodology}\label{sec:methodology}
The task of SEC is formulated as follows: given a single-channel audio input of a spoken utterance $\mathbf{X}_{l, s, e} \in\mathbb{R}^{1\times T}$, with lexical content $l$, speaker identity $s$, and annotated arousal score $e$, with its raw signal represented as a sequence of samples $x = (x_1,..., x_T)$, we design a system to synthesise $\widehat{\mathbf{Y}}_{l, s, \Bar{e}} \in\mathbb{R}^{1\times T}$, a single-channel audio that preserves the lexical content $l$ and speaker identity $s$ of $\mathbf{X}_{l, s, e}$, and converts the arousal of $\mathbf{X}_{l, s, e}$ to target arousal $\Bar{e}$.
To this, we introduce an SSL-based HiFiGAN network (see Fig.~\ref{fig:emovc_architecture}) that consists of (i) \emph{SSL encoders} that disentangles lexical $l$, speaker $s$, and emotion $e$ information, and (ii) a \emph{HiFiGAN decoder} that is trained to resynthesise emotion converted speech $\widehat{\mathbf{Y}}_{l, s, \Bar{e}}$ from the disentangled representations. The network does not rely on parallel data, where the ground-truth ${\mathbf{Y}}_{l, s, \Bar{e}}$ exists, and is generative in nature.

\subsection{Disentanglement encoders}
\noindent\textbf{Lexical Encoder:} The input to the lexical encoder $E_l$ is the time-domain signal $x$, and its output is a sequence of low frequency representations $E_l({x}) = (l_1,..., l_{T'})$. As $E_l$, we choose the pretrained SSL-based HuBERT model. Following \cite{polyak21_interspeech_resynth}, using the $k$-means algorithm, we convert the continuous HuBERT representations into discrete representations denoted as $z_l = (z_1, ..\-., z_{T'})$, where each unit $z_i$ is a positive integer and $z_i \in {0, 1, .., K}$. Existing works \cite{polyak21_interspeech_resynth, hsu2023revise}, have shown such discrete representations $z_i$ to be more related to the phonemes of the respective utterance.
\newline
\noindent\textbf{Speaker Encoder:} Similar to $E_l$, the input to the speaker encoder $E_{s}$ is the time-domain signal $x$. As $E_l$, the pretrained WavLM speaker verification model \cite{Chen2021WavLM} is used. The output of $E_{s}$ is a continuous $d-vector$ speaker representation $z_{s} \in \mathbb{R}^{512}$. Unlike the lexical representation $z_l$, $z_{s}$ is a global representation that encodes speaker information of the whole utterance. To account for the mismatch in frequency, $z_{s}$ is concatenated on all $T'$ frames of $z_l$ for the resynthesis and is denoted as $z_{T'} = (z_l, z_{s})$. 
\newline
\noindent\textbf{Emotion encoder:} To encode emotional information in the disentanglement phase, as the emotion encoder $E_e$ we use simple \textit{trainable} linear layers. The input to the encoder $E_e$ is the ground-truth annotated arousal scalar label $e \in \mathbb{R}$ and $1 \leq e \leq 7$, and the output is the emotion representation $z_{e} \in \mathbb{R}^{128}$. Similar to $z_{s}$, $z_{e}$ is also a global representation that encodes speaker information of the whole utterance $x$, and similarly $z_{e}$ is also concatenated on all frames of $z_l$ for the resynthesis and is denoted as $z_{T'} = (z_l, z_{s}, z_{e})$.

\subsection{Resynthesis decoder}
Inspired by existing works on voice conversion \cite{polyak21_interspeech_resynth, hsu2023revise} and SEC \cite{kreuk20_interspeech}, we adopt a modified version of the HiFiGAN \cite{hifigan} to resynthesise the disentangled encoder representations. Note that \cite{kreuk20_interspeech} is the closest to our work. However, in this work, we extend the HiFiGAN to non-parallel data and also tackle the thereby arising problem of disentanglement.


The HiFiGAN is trained using a generator $G$ and two discriminator networks, the multi-\textit{period} and the multi-\textit{scale} discriminators, $D_p$ and $D_s$, respectively \cite{hifigan, polyak21_interspeech_resynth}. The generator $G$ has a series of blocks composed of a transposed convolution and a dilated residual layer (see Fig.~\ref{fig:emovc_architecture}). The transposed convolutions upsample the representations to match the input number of samples $T$. The dilated residual layers increase the receptive field. The input to $G$ is the concatenated disentangled representations $z_{T'} = (z_l, z_{s}, z_{e})$ and the output is the resynthesised speech $\widehat{\mathbf{Y}}_{l, s, \Bar{e}}$ conditioned on lexical, speaker and emotion information. 

The multi-period discriminator $D_p$ consists of six sub-discriminators each operating on different \textit{period hops} of the input and generated speech: 2, 3, 4, 5, 7, and 11. Similarly, the multi-scale discriminator $D_s$ uses three sub-discriminators operating at different \textit{scales}: the original scale, x2 downsampled scale, and x4 downsampled scale.

\subsection{Loss functions}
HiFiGAN is a generative network trained using adversarial learning strategy \cite{hifigan}. Following \cite{polyak21_interspeech_resynth}, for resynthesised output speech signal $\widehat{\mathbf{Y}}$ represented as $\hat{y} = (\hat{y}_1,..., \hat{y}_T)$, each of the sub-discriminators $D_j$ in $D$ is trained to minimize the following adversarial losses (${L}_{adv}$ and ${L}_D$):
\begin{align}
    {L}_{adv} (D_j, G) &= \sum_{x} ||1 - D_j(\hat{y})||^2_2,
    \\
    {L}_D (D_j, G) &= \sum_{x} [ ||1 - D_j({x})||^2_2 + ||D_j(\hat{x})||^2_2],
\end{align}
where $x$ is the input speech and $\hat{y} = G(z_{T'}) = G(z_l, z_{s}, z_{e})$.

Along with ${L}_{adv}$ and ${L}_D$, a reconstruction loss term ${L}_{recon}$ and a feature-matching loss ${L}_{fm}$ \cite{larsen2016autoencodingFMloss} is also added to the loss function. ${L}_{recon}$ measures the Mel-spectrogram reconstruction between input $x$ and resynthesised output $\hat{y}$:
\begin{align}
    &{L}_{recon}(G) = \sum_{x} ||\phi(x) - \phi(\hat{y})||_1,
\end{align}
where $\phi$ is a function computing Mel-spectrogram. The ${L}_{fm}$ term is the distance between the discriminator activations of input $x$ and resynthesised output $\hat{y}$:
\begin{align}
    {L}_{fm}(D_j, G) = \sum_{x} \sum_{i=1}^{R} \dfrac{1}{M_i} ||\psi_i(x) - \psi_i(\hat{y})||_1,
\end{align}
where $\psi_i$ is the function that extracts the activations of the $i$-th discriminator layer, and, $M_i$ and $R$ are the number of features in $i$ and the number of layers in $D_j$, respectively.

While the concatenation of emotion label embeddings $z_{e}$ to the input of generator $G$ already conditions the resynthesised output $\hat{y}$, we also included an SER loss to the loss function. To achieve this, we use a pre-trained SSL-based SER system $E_{SER}$ introduced in \cite{wagner2023dawn}. The $E_{SER}$ network was built by fine-tuning the Wav2Vec2-Large-Robust network \cite{baevski2020wav2vec} on the MSP-Podcast (v1.7) dataset \cite{martinez2020msp}. The SER loss term ${L}_{SER}$ is formulated as,
\begin{flalign}
    {L}_{SER} = \sum_{x} [1 - {L}_{ccc}(e, E_{SER}(\hat{y}))],
\end{flalign}
where ${L}_{ccc}$ is the concordance correlation coefficient (CCC) \cite{lin_concordance_1989} that measures similarity between two variables, $e$ is the ground-truth emotion of input, and $E_{SER}(\hat{y})$ is the predicted emotion for resynthesised speech $\hat{y}$. The ${L}_{ccc}$ measure varies between $-$1 and $+$1, where $+$1 denotes perfect similarity, therefore $1 - {L}_{ccc}$ is minimized during training.

The final loss for generator $G$ and discriminator $D$ is:
\begin{align}
    {L}_{G}(D, G) = &\sum_{j=1}^{J} [{L}_{adv} (D_j, G) + \lambda_{fm}{L}_{fm}(D_j, G)] 
    \nonumber\\ &+ \lambda_{r}{L}_{recon}(G) +  \lambda_{SER}{L}_{SER}, \\
    {L}_{D}(D, G) = &\sum_{j=1}^{J} {L}_D (D_j, G),
\end{align}
where $J$ is the number of sub-discriminators in $D$. Following \cite{polyak21_interspeech_resynth}, we set $\lambda_{fm} = 2$ and $\lambda_{r} = 45$. $\lambda_{SER}$ was set to 1 after preliminary results supported this setup.


\begin{table}[t!]
        \begin{tabular}{lc|cc}
        \toprule
        \multirow{2}{*}{Model versions} & \multirow{2}{*}{WVMOS} & \multicolumn{2}{c}{SER Error}                         \\
                                          &                            & $L_{mse}$                        & $L_{abs}$                        \\ \midrule
        $z_l$              & 1.80        & $-$         & $-$             \\
        $z_l + z_{s}$    & 2.29       & $-$    & $-$                     \\ \midrule
        $z_l + z_{s} + z_{e}$ &   2.64    & 0.0971    & 0.2642   \\
        $z_l + z_{s} + z_{e} + L_{SER}$   & \textbf{3.26}$^*$  & \textbf{0.0843}$^*$ &   \textbf{0.2442}$^*$ \\ \bottomrule
        \end{tabular}
    \caption{Overall performance of model versions. * indicates statistically significant improvements in results.}
    \label{tab:overall_compare}
\end{table}

\section{Experimental Setup}\label{sec:exp-setup}
\noindent\textbf{Dataset:} The dataset used in this work is the \emph{in-the-wild} MSP-Podcast dataset (v1.10) \cite{martinez2020msp}. To the best of our knowledge, this is the first work in literature that performs SEC on an in-the-wild dataset. The dataset consists of approximately 166hrs of audio collected from podcasts. Emotion is labeled at the utterance-level in terms of arousal, valence, and dominance. In this work, we only use the arousal annotations for SEC. The arousal annotations, collected on a scale of 1 to 7, are distributed with $\mu=$4 and $\sigma=$0.95, denoting that there are comparatively fewer samples on the extremes of the scales (1 and 7) than on middle regions (3 to 5). Note this also reflects the nature of emotions in-the-wild podcasts and also in any real-world scenario, where extreme emotions are expected to be sparse. 
\\
\noindent\textbf{Validation measures:} We validate the proposed methodology in terms of both the SEC capabilities and the naturalness of synthesised speech. As the measure of SEC capability, we use the mean-squared ${L}_{mse}$ and mean-absolute ${L}_{abs}$ errors, calculated between the target arousal $\Bar{e}$ and the SER prediction on the resynthesised output $E_{SER}(\hat{y})$. As the measure of the naturalness of $\hat{y}$, we use the wav2vec mean-opinion score (WVMOS) \cite{wvmos}. WVMOS is an objective speech quality measure based on wav2vec2.0 \cite{baevski2020wav2vec} and is fine-tuned on the mean-opinion scores (on a scale of 1 to 5) obtained from the listening tests of the 2018 Voice Conversion Challenge \cite{vcChallenge2018}. As the listening tests primarily focused on the naturalness of the voice conversion task, it makes the WVMOS well suited for non-intrusively measuring the naturalness of $\hat{y}$. Note here that, as we do not use parallel data we can only rely on non-intrusive measures that do not require the ground-truth audio of emotion conversion $\mathbf{Y}_{l, s, \Bar{e}}$. Statistical significance for improved performance is estimated using one-tailed $t$-test on error distributions, asserting significance for \emph{p}-values $\leq0.05$.

\section{Results and Discussion}\label{sec:results}
\noindent\textbf{Overall performance: } Firstly, we compare the overall performance of four different versions of the proposed met\-hodology: (i) $z_l$, which only uses the lexical representations for resynthesis, (ii) $z_l + z_s$, which also uses the speaker information $z_s$, (iii) $z_l + z_s + z_e$, which also uses the emotion label embeddings $z_e$ thereby also conditioning on emotion information, and (iv) $z_l + z_s + z_e + L_{SER}$ which further conditions on emotion information by including $L_{SER}$ in the loss function. As this is the first work in literature to perform SEC in terms of \textit{arousal}, we do not have any baselines, rather we present different versions of the proposed methodology for comparison.

From the results presented in Table~\ref{tab:overall_compare}, we note the following: firstly, with the increasing addition of disentangled representations the HiFiGAN is capable of producing more natural-sounding speech. The usage of all three representations along with the inclusion of SER loss $z_l + z_s + z_e + L_{SER}$ achieves the best WVMOS score of $3.26$ with \textit{statistical significance}. This also validates the SEC capability of the proposed approach, as conditioning on emotion should also improve the naturalness of synthesised speech and it is reflected in the WVMOS score. In \cite{polyak21_interspeech_resynth}, it is suggested to also include F0 representations to obtain better natural-sounding speech. However, in our case, we do not use F0 representations for resynthesis as conditioning on F0 will also affect the conditioning on emotion representations $z_e$, given that F0 and emotion are highly correlated \cite{egemaps}. 
Secondly, including the SER loss term $L_{SER}$ for further conditioning on emotional content results in an improved SEC performance. The model achieves an $L_{mse}$ of $0.0843$ and $L_{abs}$ of $0.2442$, improving over $z_l + z_s + z_e$ with \textit{statistical significance}. The SEC capabilities can be further noted in the audio examples presented online\footnote{\url{https://uhh.de/inf-sp-emovc-itg} \label{link:audio_sample}}. 


\begin{figure}[t!]
\centering
    \begin{subfigure}{0.23\textwidth}
        \centering
    \includegraphics[width=\textwidth]{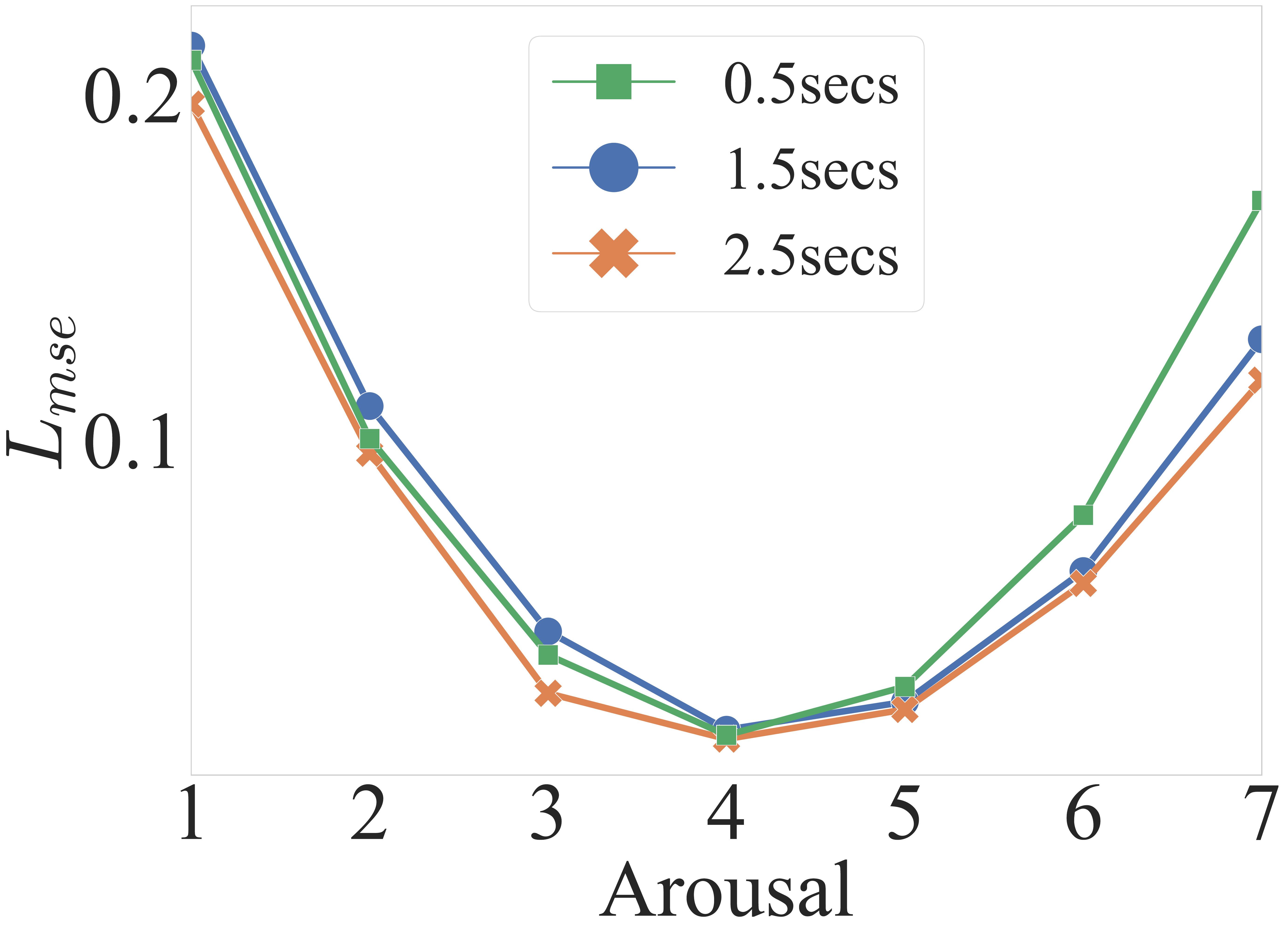}
        \subcaption{$L_{mse}$ trends}
        \label{fig:serloss_trends}
    \end{subfigure}
\centering
    \begin{subfigure}{0.23\textwidth}
        \centering
        \includegraphics[width=\textwidth]{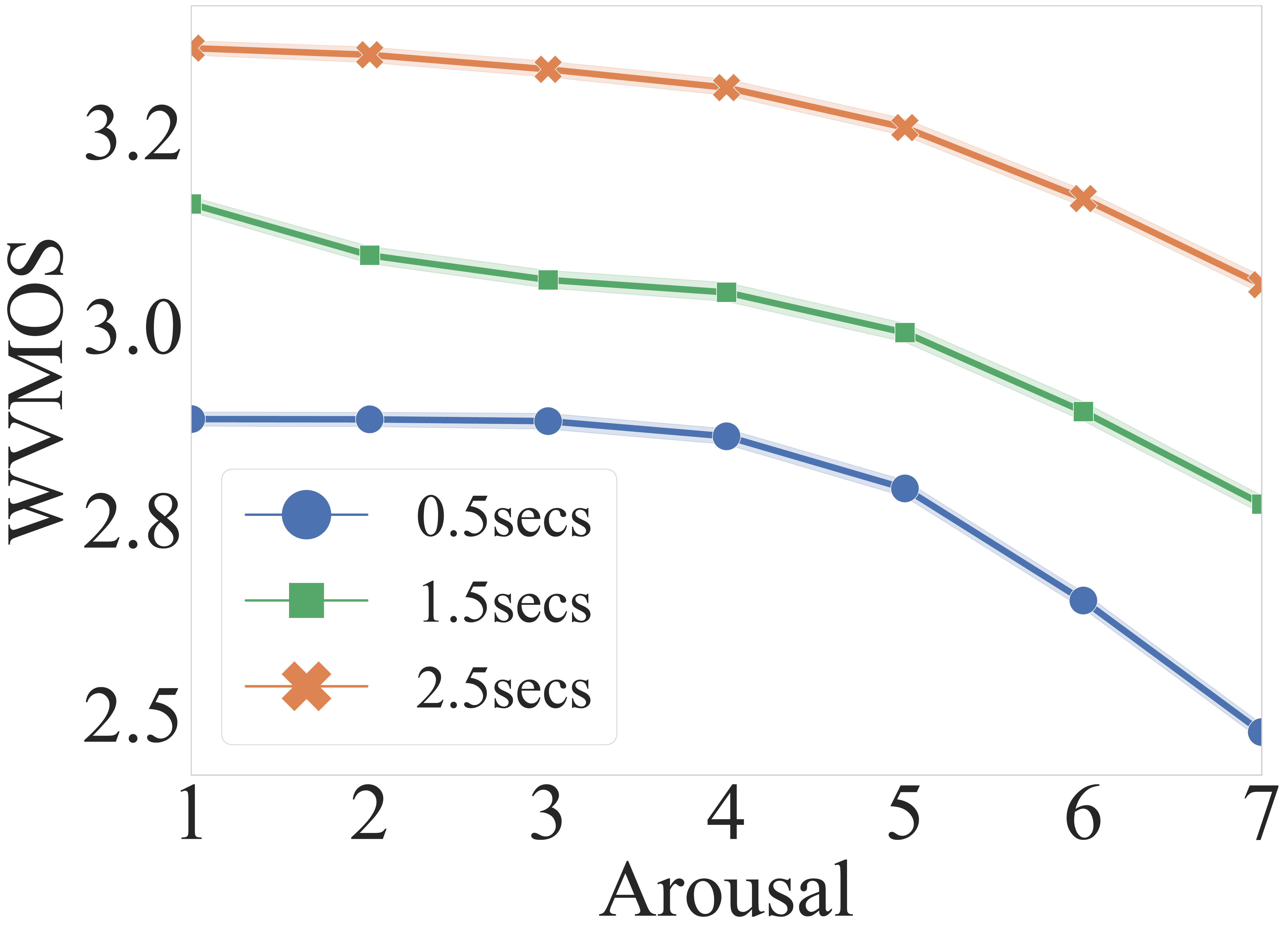}
        \subcaption{WVMOS trends}
        \label{fig:wvmos_trends}
    \end{subfigure}
    \caption{Class-wise performance for target arousal $\Bar{e}$, with respect to the input segment sizes.}
    \label{fig:arousal_levelwise_trends}
\end{figure}

\noindent\textbf{Influence of segment size: } While training HiFiGAN, to account for varying utterance lengths, segments of consistent lengths are randomly sampled from different regions of the full utterance and then are used for training. In \cite{polyak21_interspeech_resynth}, it is noted that the HiFiGAN is robust enough to resynthesise natural-sounding speech even when trained on a small segment size (0.75s). However, for modeling emotion the context is important \cite{ramet2018context}, thereby requiring larger segment sizes to aptly condition on the emotion content. 

To investigate the influence of segment sizes on SEC, we trained the proposed methodology on varying segment sizes, and the results are presented in Fig.~\ref{fig:arousal_levelwise_trends}. From the results, it is noted that larger segment sizes improve SEC performances, both in terms of $L_{mse}$ and WVMOS. While the improvements on naturalness (WVMOS) are clearly noted, improvements on $L_{mse}$ are larger for extreme emotion values (1 and 7) than mid-ranges of emotion (3 to 5).

\noindent\textbf{Performance on different arousal classes: } Existing research has shown that SEC techniques generally tend to perform well on certain emotion pairs than others. For example, in \cite{rizos2020stargan}, it is noted that the emotion pair of angry-sad is easier to convert than the happy-angry pair. To investigate this, Fig.~\ref{fig:arousal_levelwise_trends} also present performances with respect to each of the arousal classes ranging from 1 to 7. In terms of the SEC capability, the $L_{mse}$ error is noted to be higher for extreme arousal classes (1 and 7). However, for mid-scale arousal values between 2 and 6 the $L_{mse}$ error is comparatively smaller. In terms of the naturalness of synthesised speech, the WVMOS score is lower for high arousal classes (6 and 7) and is comparatively better for other arousal classes (1 to 5). This reveals that the proposed methodology is capable of synthesising more natural-sounding speech when \textit{reducing} the arousal of the input speech, than when \textit{increasing} the arousal. Overall, the results reveal that the methodology better synthesises speech for mid-scale arousal than for extreme arousal.

\begin{figure}[t!]
\centering
\includegraphics[width=\columnwidth]{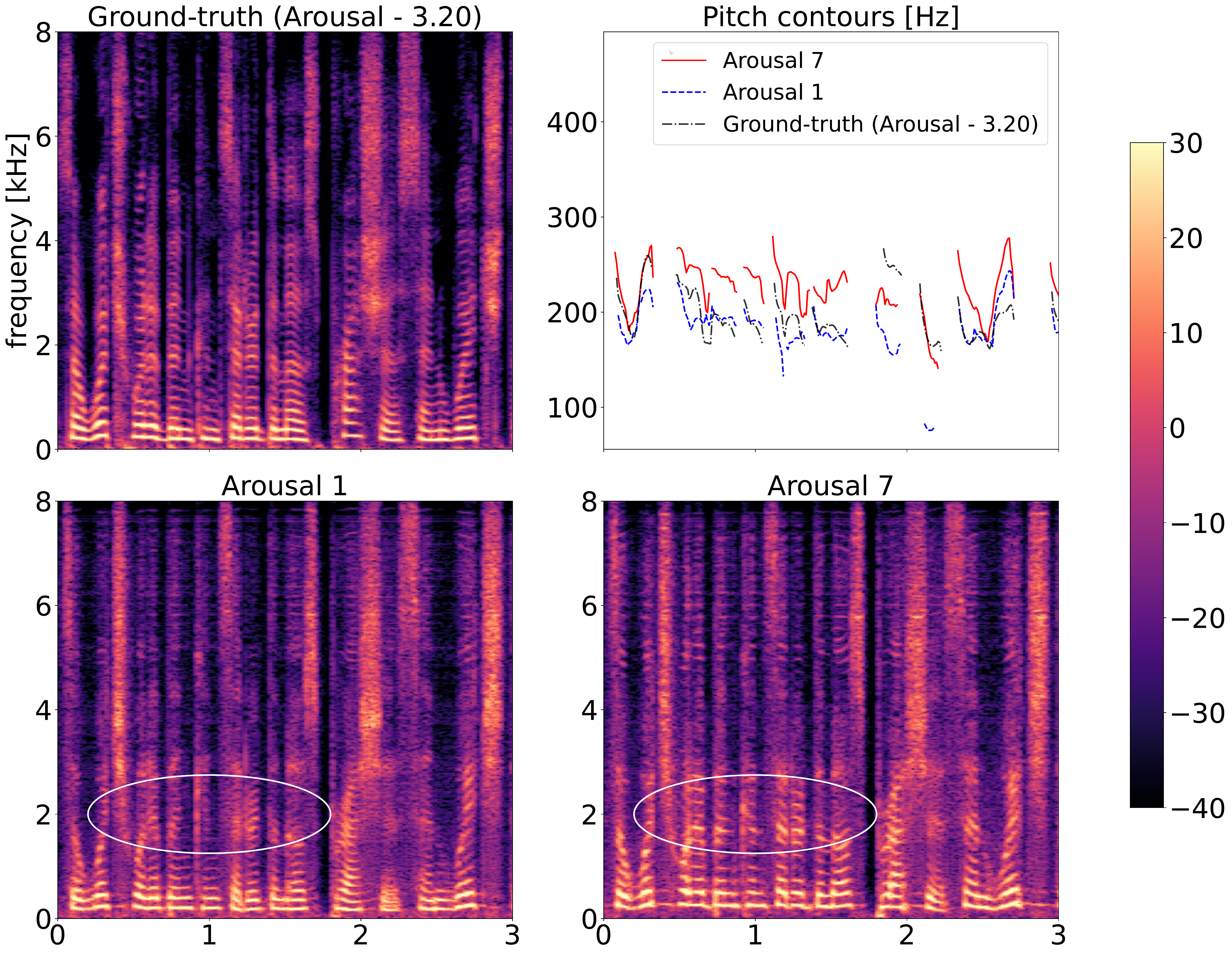}
	\caption{Sample \text{log}-energy spectrogram of emotion converted speech, along with comparisons on pitch contours.}
	\label{fig:sample_spec}%
\end{figure} 

\noindent\textbf{Qualitative analysis of spectrograms: } Sample spectrograms of emotion converted speech can be seen in Fig.~\ref{fig:sample_spec}, for the ground-truth speech of arousal $e=3.20$, for synthesised speech of \textit{reduced} arousal $\Bar{e}=1$, and for \textit{increased} arousal $\Bar{e}=7$. Comparing the spectrograms of arousal $1$ and arousal $7$, from the marked eclipses, we can observe that for increased arousal of $7$ the spectrograms have larger magnitudes in the mid-frequencies. This reveals that the model associates larger frequency magnitudes for high arousal speech, than for low arousal speech. Along with the spectrograms, we also plot the pitch contours of the respective speech signals. From the pitch contours, it can be noted that the synthesised speech for high arousal ($\Bar{e}=7$) has a higher mean and variability of pitch, than that of both the ground-truth speech ($e=3.20$) and the synthesised speech for low arousal ($\Bar{e}=1$). This aligns with existing literature that associates high intensity of emotion with an increased mean pitch \cite{zhou2023_emointensitycontrol}. This difference in pitch is also clearly notable in the audio examples available online\textsuperscript{\ref{link:audio_sample}}. These results validate that the proposed model successfully performs SEC by aptly conditioning on the emotion content.


\section{Conclusion}\label{sec:conclusion}
In this paper, as the first in literature, we tackle the problem of in-the-wild speech emotion conversion on continuous arousal representations, as opposed to acted speech and categorical representations. 
In-the-wild datasets lack parallel utterances and thereby the problem of disentangling the lexical, speaker, and emotion information arises. 
To tackle this, we introduced a novel methodology that uses SSL encoders for disentanglement, and a HiFiGAN vocoder to resynthesise emotion conditioned speech from the disentangled SSL representations. We validated the network on an in-the-wild dataset, in terms of its emotion conversion capability, using a pretrained SER system, and the naturalness of synthesised speech, using WVMOS. Results reveal that the network is capable of synthesising natural-sounding speech with emotion conversion. Further analysis revealed that the network better synthesises emotional speech for mid-scale arousal than for extreme arousal. Finally, the pitch contour analysis showed that the synthesised speech for high arousal has a higher mean and variability of pitch than that of both the ground-truth speech and the synthesised speech for low arousal.



\newpage

\small
\bibliographystyle{ieeetr}
\bibliography{refs}

\def\ICASSP{IEEE Int. Conf. on Acoustics, Speech, Sig. Proc.,
  ICASSP}\def\INTERSPEECH{Interspeech}\def\ACII{IEEE Int. Conf. on Affective
  Comp. and Intelligent Interaction}\def\NIPS{Advances in Neural Inf. Proc.
  Sys., NeurIPS}\def\ICML{Int. Conf. Machine Learning (ICML)}\def\TAC{IEEE
  Tran. on Affective Computing}\def\TASL{IEEE Tran. on on Audio, Speech, and
  Language Processing}
\begin{thebibliography}{10}

\bibitem{schuller2018speech}
B.~W. Schuller, ``Speech emotion recognition: Two decades in a nutshell,
  benchmarks, and ongoing trends,'' {\em Communications of the ACM}, vol.~61,
  no.~5, pp.~90--99, 2018.

\bibitem{crumpton2016survey}
J.~Crumpton and C.~L. Bethel, ``A survey of using vocal prosody to convey
  emotion in robot speech,'' {\em International Journal of Social Robotics},
  vol.~8, pp.~271--285, 2016.

\bibitem{zhou2022mixedemo}
K.~Zhou, B.~Sisman, R.~Rana, B.~W. Schuller, and H.~Li, ``Speech synthesis with
  mixed emotions,'' {\em \TAC}, 2022.

\bibitem{hifigan}
J.~Kong, J.~Kim, and J.~Bae, ``Hifi-gan: Generative adversarial networks for
  efficient and high fidelity speech synthesis,'' {\em \NIPS}, vol.~33, 2020.

\bibitem{hsu2023revise}
W.-N. Hsu, T.~Remez, B.~Shi, J.~Donley, and Y.~Adi, ``Revise: Self-supervised
  speech resynthesis with visual input for universal and generalized speech
  regeneration,'' in {\em Proc. of the IEEE/CVF Conf. on Computer Vision and
  Pattern Recognition}, 2023.

\bibitem{triantafyllopoulos2023overview}
A.~Triantafyllopoulos, B.~W. Schuller, G.~{\.I}ymen, M.~Sezgin, X.~He, Z.~Yang,
  P.~Tzirakis, S.~Liu, S.~Mertes, E.~Andr{\'e}, {\em et~al.}, ``An overview of
  affective speech synthesis and conversion in the deep learning era,'' {\em
  Proc. of the IEEE}, 2023.

\bibitem{zhou2022emotional}
K.~Zhou, B.~Sisman, R.~Liu, and H.~Li, ``Emotional voice conversion: Theory,
  databases and esd,'' {\em Speech Communication}, vol.~137, pp.~1--18, 2022.

\bibitem{VCdu22c_interspeech}
Z.~Du, B.~Sisman, K.~Zhou, and H.~Li, ``{Disentanglement of Emotional Style and
  Speaker Identity for Expressive Voice Conversion},'' in {\em \INTERSPEECH},
  2022.

\bibitem{ekman1971constants}
P.~Ekman and W.~V. Friesen, ``{Constants across cultures in the face and
  emotion.},'' {\em Journal of personality and social psychology}, vol.~17,
  no.~2, p.~124, 1971.

\bibitem{busso2008iemocap}
C.~Busso, M.~Bulut, C.-C. Lee, A.~Kazemzadeh, E.~Mower, S.~Kim, J.~N. Chang,
  S.~Lee, and S.~S. Narayanan, ``Iemocap: Interactive emotional dyadic motion
  capture database,'' {\em Language resources and evaluation}, vol.~42, no.~4,
  pp.~335--359, 2008.

\bibitem{russell1980circumplex}
J.~A. Russell, ``A circumplex model of affect.,'' {\em Journal of personality
  and social psychology}, vol.~39, no.~6, p.~1161, 1980.

\bibitem{martinez2020msp}
L.~Martinez-Lucas, M.~Abdelwahab, and C.~Busso, ``The msp-conversation
  corpus,'' {\em \INTERSPEECH}, 2020.

\bibitem{prabhu22_interspeech}
N.~Raj~Prabhu, G.~Carbajal, N.~Lehmann-Willenbrock, and T.~Gerkmann,
  ``End-to-end label uncertainty modeling for speech-based arousal recognition
  using \text{Bayesian} neural networks,'' in {\em \INTERSPEECH}, Sep 2022.

\bibitem{zhou2023_emointensitycontrol}
K.~Zhou, B.~Sisman, R.~Rana, B.~W. Schuller, and H.~Li, ``Emotion intensity and
  its control for emotional voice conversion,'' {\em \TAC}, vol.~14, no.~1,
  pp.~31--48, 2023.

\bibitem{vawganzhou2021seen}
K.~Zhou, B.~Sisman, R.~Liu, and H.~Li, ``Seen and unseen emotional style
  transfer for voice conversion with a new emotional speech dataset,'' in {\em
  \ICASSP}, pp.~920--924, IEEE, 2021.

\bibitem{burkhardt2005database}
F.~Burkhardt, A.~Paeschke, M.~Rolfes, W.~F. Sendlmeier, B.~Weiss, {\em et~al.},
  ``A database of german emotional speech.,'' in {\em \INTERSPEECH}, vol.~5,
  pp.~1517--1520, 2005.

\bibitem{rizos2020stargan}
G.~Rizos, A.~Baird, M.~Elliott, and B.~Schuller, ``Stargan for emotional speech
  conversion: Validated by data augmentation of end-to-end emotion
  recognition,'' in {\em \ICASSP}, 2020.

\bibitem{zhour2020CycleGAN}
K.~Zhou, B.~Sisman, and H.~Li, ``Transforming spectrum and prosody for
  emotional voice conversion with non-parallel training data,'' in {\em The
  Speaker and Language Recognition Workshop (Speaker Odyssey)}, 05 2020.

\bibitem{vawgan_base}
K.~Zhou, B.~Sisman, and H.~Li, ``Vaw-gan for disentanglement and recomposition
  of emotional elements in speech,'' in {\em IEEE Spoken Language Technology
  Workshop}, 2021.

\bibitem{deoliveira2023leveraging}
D.~de~Oliveira, N.~Raj~Prabhu, and T.~Gerkmann, ``Leveraging semantic
  information for efficient self-supervised emotion recognition with
  audio-textual distilled models,'' in {\em \INTERSPEECH}, 2023.

\bibitem{wagner2023dawn}
J.~Wagner, A.~Triantafyllopoulos, H.~Wierstorf, M.~Schmitt, F.~Burkhardt,
  F.~Eyben, and B.~W. Schuller, ``Dawn of the transformer era in speech emotion
  recognition: closing the valence gap,'' {\em IEEE Tran. on Pattern Analysis
  and Machine Intelligence}, 2023.

\bibitem{polyak21_interspeech_resynth}
A.~Polyak, Y.~Adi, J.~Copet, E.~Kharitonov, K.~Lakhotia, W.-N. Hsu, A.~Mohamed,
  and E.~Dupoux, ``{Speech Resynthesis from Discrete Disentangled
  Self-Supervised Representations},'' in {\em \INTERSPEECH}, 2021.

\bibitem{baevski2020wav2vec}
A.~Baevski, Y.~Zhou, A.~Mohamed, and M.~Auli, ``wav2vec 2.0: A framework for
  self-supervised learning of speech representations,'' {\em \NIPS}, vol.~33,
  pp.~12449--12460, 2020.

\bibitem{kreuk20_interspeech}
F.~Kreuk, J.~Keshet, and Y.~Adi, ``{Self-Supervised Contrastive Learning for
  Unsupervised Phoneme Segmentation},'' in {\em \INTERSPEECH}, pp.~3700--3704,
  2020.

\bibitem{Chen2021WavLM}
S.~Chen, C.~Wang, Z.~Chen, Y.~Wu, S.~Liu, Z.~Chen, J.~Li, N.~Kanda,
  T.~Yoshioka, X.~Xiao, J.~Wu, L.~Zhou, S.~Ren, Y.~Qian, Y.~Qian, M.~Zeng, and
  F.~Wei, ``Wavlm: Large-scale self-supervised pre-training for full stack
  speech processing,'' {\em IEEE Journal of Selected Topics in Signal
  Processing}, vol.~16, pp.~1505--1518, 2021.

\bibitem{dischubertAnalysis2023}
A.~Sicherman and Y.~Adi, ``Analysing discrete self supervised speech
  representation for spoken language modeling,'' in {\em \ICASSP}, pp.~1--5,
  2023.

\bibitem{wavnet}
A.~van~den Oord, S.~Dieleman, H.~Zen, K.~Simonyan, O.~Vinyals, A.~Graves,
  N.~Kalchbrenner, A.~Senior, and K.~Kavukcuoglu, ``Wavenet: A generative model
  for raw audio,'' in {\em 9th ISCA Speech Synthesis Workshop}, pp.~125--125,
  2016.

\bibitem{larsen2016autoencodingFMloss}
A.~B.~L. Larsen, S.~K. S{\o}nderby, H.~Larochelle, and O.~Winther,
  ``Autoencoding beyond pixels using a learned similarity metric,'' in {\em
  \ICML}, 2016.

\bibitem{lin_concordance_1989}
L.~I.-K. Lin, ``A {Concordance} {Correlation} {Coefficient} to {Evaluate}
  {Reproducibility},'' {\em Biometrics}, vol.~45, p.~255, Mar. 1989.

\bibitem{wvmos}
P.~Andreev, A.~Alanov, O.~Ivanov, and D.~Vetrov, ``Hifi++: A unified framework
  for bandwidth extension and speech enhancement,'' in {\em \ICASSP}, pp.~1--5,
  2023.

\bibitem{vcChallenge2018}
J.~Lorenzo-Trueba, J.~Yamagishi, T.~Toda, D.~Saito, F.~Villavicencio,
  T.~Kinnunen, and Z.-H. Ling, ``The voice conversion challenge 2018: Promoting
  development of parallel and nonparallel methods,'' in {\em \INTERSPEECH}, Apr
  2018.

\bibitem{egemaps}
F.~Eyben, K.~R. Scherer, B.~W. Schuller, J.~Sundberg, E.~Andre, C.~Busso, L.~Y.
  Devillers, J.~Epps, P.~Laukka, S.~S. Narayanan, and K.~P. Truong, ``The
  geneva minimalistic acoustic parameter set (gemaps) for voice research and
  affective computing,'' {\em \TAC}, vol.~7, pp.~190--202, Apr 2016.

\bibitem{ramet2018context}
G.~Ramet, P.~N. Garner, M.~Baeriswyl, and A.~Lazaridis, ``Context-aware
  attention mechanism for speech emotion recognition,'' in {\em IEEE Spoken
  Language Technology Workshop}, pp.~126--131, 2018.

\end{thebibliography}


\end{document}